# Polarization spontaneous and piezo: fundamentals and their implementation in *ab initio* calculations

Pawel Strak[1], Pawel Kempisty[1,2], Konrad Sakowski[1,3], Jacek Piechota[1], Izabella Grzegory[1], Eva Monroy[5], Agata Kaminska[1,4,6] and Stanislaw Krukowski[1]*

[1]*Institute of High Pressure Physics, Polish Academy of Sciences, Sokolowska 29/37, 01-142 Warsaw, Poland*

[2]*Research Institute for Applied Mechanics, Kyushu University, Fukuoka 816-8580, Japan*

[3]*Institute of Applied Mathematics and Mechanics, University of Warsaw, 02-097 Warsaw, Poland*

[4]*Institute of Physics, Polish Academy of Sciences, Aleja Lotnikow 32/46, PL-02668 Warsaw, Poland*

[5]*Univ. Grenoble-Alpes, CEA, Grenoble INP, IRIG, PHELIQS , 17 av. des Martyrs, 38000 Grenoble, France*

[6]*Cardinal Stefan Wyszynski University, Faculty of Mathematics and Natural Sciences. School of Exact Sciences, Dewajtis 5, 01-815 Warsaw, Poland*

Abstract

Fundamental properties of spontaneous and piezo polarization are reformulated and critically reviewed. It was demonstrated that Landau definition of polarization as a dipole density could be used to the infinite systems. The difference between the bulk polarization and surface polarity are distinguished thus creating clear identification of both components. The local model of spontaneous polarization was created and used to calculate spontaneous polarization as the electric dipole density. The proposed local model correctly predicts c-axis spontaneous polarization values of the nitride wurtzite semiconductors. In addition, the model results are in accordance with polarization equal to zero for zinc blende lattice. The spontaneous polarization values obtained for all wurtzite III nitrides are in basic agreement with the earlier calculations using Berry phase. Wurtzite nitride superlattices *ab initio* calculations were performed to derive polarization-induced fields in the coherently strained lattices showing good agreement with the polarization values. The strained superlattice data were used to determine the piezoelectric parameters of wurtzite nitrides obtaining the values that were in basic agreement with the earlier data. Zinc blende superlattices were also modeled using *ab initio* calculations showing results that are in agreement with the absence of polarization of all nitrides in zinc blende symmetry.

*Corresponding author, email: stach@unipress.waw.pl

**1. Introduction.**

Polarization is an important macroscopic vectorial quantity emerging in the systems having symmetry groups that allow the system to attain its nonzero values [1]. Spontaneous polarization is the specific aspect of this phenomenon in which the system attains the state of nonzero polarization without any inference from the outside. The phenomenon is an inherent property of the system and therefore it is defined in isolation. On the other hand, it is relatively easy to affect the state of the system and induce polarization by mere application of the electric field from the outside. The field breaks the system symmetry leading to the polarization. Therefore, the determination of the polarization in general and its spontaneous variation requires a precise definition of the external conditions. Paradoxically, in some cases, manipulation of the system from outside is supportive for the determination of the spontaneous polarization, despite the fact that its definition assumes no such influence.

The most prominent group of standard semiconductors are those having wurtzite and zinc blende lattices. Despite noticeable/apparent similarities, in some respect, they are drastically different. Wurtzite crystalline symmetry allows the occurrence of polarization while zinc blende does not. Macroscopically, polarization occurs due to the relative shift of the center of the negative electron charge with respect to the position of the positive atomic core, i.e. creation of electric dipole density [1]. This interpretation may be also applied to finite-size systems, such as molecules or nanoobjects [2].

Polarization effects are important, they affect the physical properties of semiconductor systems through the emergence of electric fields of various magnitudes and ranges. In large-size systems, the macroscopic electric fields are negligible due to charge screening, known as Debye-Hückel or Thomas-Fermi effects [3,4]. A much stronger influence of polarization-induced electric fields is observed in nanometer-scale systems. A glaring, positive example of polarization application is localization of electrons by the electric field in the GaN-based field-effect transistors (FETs) [5,6]. In laser diodes (LDs) and light-emitting diodes (LEDs), based on III-nitride multi-quantum-wells (MQWs), the polarization related electric fields are highly detrimental, reducing electron-hole wavefunction overlap and consequently the radiative recombination efficiency by the so called quantum-confined Stark effect (QCSE) [7-11]. In devices containing heterostructures, polarization difference entails a sheet charge [5,6,12,13] and a surface dipole layer [21,35] at the heterointerfaces.

Polarization was defined first by Nobel Prize winner L. D. Landau as electric dipole density, i.e. the magnitude of the electric dipole for the unit of the volume or for the separated molecule [1]. While in the case of the molecule, its finite size did not cause any fundamental problems, in the case of the infinite solids polarization as a bulk property was questioned. At the beginning, Martin claimed that the property cannot be obtained from unit cell calculation because of the charge transfer between various cells and contribution from the surface states [14]. Posternak et al. calculated polarization of BeO showing that the spontaneous polarization was not accessible in the procedure using periodic boundary conditions (PBC) [15]. Accordingly, Tagantsev claimed that there is no possibility to define spontaneous polarization as a bulk property [16]. In a series of later papers, Springborg, Kirtman et al. showed that the polarization as bulk property is critically affected by the edge termination which cannot be removed by extending the size of the system to infinity [17-21]. They concluded that the polarization as the bulk property cannot be uniquely determined as the result depends on the boundaries and also on the shape of simulated system [21]. This argument was also used by Spaldin who showed that, depending on the termination, the two different values of polarization could be obtained for different termination within a simple Clausius-Mosotti model in which continuous charge distribution is replaced by a set of positively and negatively charged ions [22]. In the case of continuous electron charge distribution, this translates into an infinite number of the polarization values.

As a remedy, a different approach was developed in which the polarization change was calculated [23-26]. The idea was proposed first by King and Vanderbilt who declared that the polarization is equivalent to surface charge density but modulo the charge unit $e/A_{surf}$ or $2e/A_{surf}$ [23,24]. This was in agreement with the results obtained by Resta, who declared that polarization calculated as dipole density is ill-defined quantity [25,26]. Generally, this is a correct statement as the calculated dipole depends on the selection of the cell boundaries, which undergoes the jump when the shifted cell boundary is crossed by the atom. Nevertheless, Resta declared that the polarization is intrinsically bulk property, which is basically in agreement with Landau's statement. To resolve the problem he proposed to calculate the polarization change expressed in terms of geometrical phase or Berry phase related to polarization current induced during the change of the polarization of the system due to predefined transformation. It has to be stressed that Resta derived this expression from Landau's definition [25]. In his derivation, the polarization definition of Landau was transformed into the calculation of polarization current expressed in terms of Wannier functions [26]. The polarization change was determined as bulk quantity that can be determined using a periodic unit cell. The obtained property was claimed by Resta to be the only valid polarization as a bulk quantity.

Using this definition, Fiorentini et al. calculated the polarization change for wurtzite AlN, GaN and InN using zinc blende lattice as a reference, i.e. assuming zinc blende polarization is zero [28,28]. As expected, the spontaneous polarization in the zinc blende lattice is supposed to vanish by symmetry, therefore the obtained polarization change values are considered as the total spontaneous polarization of the wurtzite nitrides. Much later Dreyer et al. used the same procedure [29]. The difference was that the spontaneous polarization difference was calculated between wurtzite and the artificially designed hexagonal phase [29]. The latter has zero polarization due to mirror symmetry with respect to the $xy$ plane. The results of Fiorentini et al. and Dreyer et al. are drastically different, the latter's polarization values are more than one order of magnitude higher. Dreyer et al. claimed that the difference is due to the fact that polarization in zinc blende is not zero. Recently, Yoo et al calculated the spontaneous polarization of wurtzite and zinc blende GaN and AlN [12]. They obtained the wurtzite polarization values in agreement with the earlier results of Dreyer et al. In addition, nonzero polarization values of zinc blende GaN and AlN were listed. The latter nonzero value could be claimed, nevertheless, the spontaneous polarization of the wurtzite should be independent from the reference value, provided that the sound procedure is used. If so, then from the obtained polarization difference, it follows that the polarization of the zinc blende is comparable to that of the wurtzite. This is definitely not true, therefore this difference has to be explained.

In addition to the Berry phase direct approaches, the indirect route was used, based on determination of electric fields from *ab initio* models employed in the modelling of polar quantum wells and polar surfaces. *Ab initio* calculations were used for simulations of the multiquantum wells (MQWs) that form active layers of light-emitting diodes (LEDs) and laser diodes (LDs) [9-11]. Naturally, as an example for calculation, AlGaN or GaInN solid solutions-based wells and barriers are poor candidates, therefore the polar simple GaN/AlN MQWs were considered. These structures have their properties affected by the polarization-induced field along 0z axis. In such structures, embedded in the external solid the electric fields emerge, due to barrier-well polarization difference [8,9,30]. This is perfectly simulated by *ab initio* calculation of a single AlN/GaN period because the total potential difference across the well/barrier structure is zero [8-11]. Such calculations were made for the ideal wurtzite and zinc blende lattice, in which Al and Ga atoms are located in the ideal lattices, having either GaN or AlN lattice parameters [11]. The results proved that electric field arises in wurtzite lattice but it is zero in zinc blende lattice. Naturally, these results are obtained within the precision of the potential averaging and finite system size, nevertheless, it is estimated that the fields in zinc blende are at least two orders of magnitude lower. These results do not prove that the zinc blende has zero polarization, merely that the GaN and AlN polarizations are identical in the zinc blende ideal lattices strained either to GaN or to AlN. The relaxation of the lattice leads to the emergence of the fields because the ideal zinc blende symmetry is broken by the strain. In summary it is strong indication only that the spontaneous polarization in zinc blende is zero.

Another indirect approach was based on slab simulations used for surface modeling. Spontaneous polarization is defined as polarization of the solid in the absence of external electric charge [31]. As discussed by Boguslawski and Bernholz, this is equivalent to the zero electric displacement field in the entire system. This leads to the existence of the polarization-induced electric field in the sample. Application of the external field could compensate this field to zero [11]. The nitride slab with no charged surface states at the specially formed boundaries was subject to the external field to obtain zero fields inside. From the magnitude of the applied external field, the spontaneous polarization was deduced. Still, the relation between the Berry phase and slab results requires explanation. These simulations provided different values of polarization. These results need to be further verified as the slab contains fractionally charged surface states which could be additionally charged due to the external field, affecting the relation between the external field and the polarization. An attempt to compare these differences was undertaken recently [32]. The investigations included the influence of the piezo effect. Despite the large discrepancy of the magnitude of polarization, the differences are similar and could be easily compensated by the strain induced piezo effects. Therefore no conclusive determination was possible, the results was that all these values sets are possible. Thus the polarization in the infinite solids is not determined precisely.

A completely different status was achieved in the studies of polarization of finite objects, i.e. molecules, nanoclusters [2]. Polarization of the finite objects is defined as a total magnetic moment that is calculated using various formulations and also measured experimentally [33]. The electric dipole moments are calculated and compared using large number of various numerical methods [34]. Good accuracy was achieved with the errors of the order of a few percent. Therefore the problem of polarization of the finite objects is solved.

The problem of the spontaneous polarization in the infinite solids is not solved. The present paper is devoted to resolving the difficulties and providing definitive answers to these questions. Therefore we define the spontaneous polarization first and discuss the difference between bulk polarization and the polar surface effects. Then the new local calculation method of the spontaneous polarization is presented. This will be described in Section 2 devoted to the basic model. As the present state of the field definitely requires basic formulation, this Section, preceding the presentation of the calculation method, is introduced. Then the results obtained for the nitrides: BN, AlN, GaN and InN are presented. Both the local bulk model and the supercell data are discussed. Finally, the present results are critically compared to the previously obtained data.

## 2. The model

Spontaneous polarization is a bulk property that leads to the electric field which affects the properties of a semiconductor. The emerging field in the flat uniform slab does not depend on its thickness, i.e. it is equivalent to the electric capacitor. Accordingly, the identical electric potential can be obtained assuming electric charge density on both polar surfaces. An identical type of contribution stems for the charged surface effect, therefore these effects are intermingled.

This scenario was applied in the polarization theory by Spaldin [22]. In a tutorial approach, she demonstrated the application of the modern theory of polarization in experimental determination by Sawyer-Tower method [22]. In particular, in the diagram in Fig. 1 of Ref 11 the author presents the two representative unit cells that could lead to the two opposite signs of polarization values. This is further confirmed by the Fig. 2 in which the author presents the two polarizations represented by the surface charges. This interpretation is compatible with the original arguments of King-Smith and Vanderbilt [23,24] and also by Ambacher et al. [35] that the polarization could be represented by surface charge. It is easy to show that this picture leads to the oversimplified interpretation of the phenomenon.

In order to prove this we would follow the Spaldin argument in application to the wurtzite and zinc blende lattices. We applied the simple model to zinc blende GaN and wurtzite GaN slabs terminated by $GaN(11 \pm 1)$ and $GaN(000 \pm 1)$ polar surfaces as presented in Figs. 1 and 2, for zinc blende and wurtzite respectively.

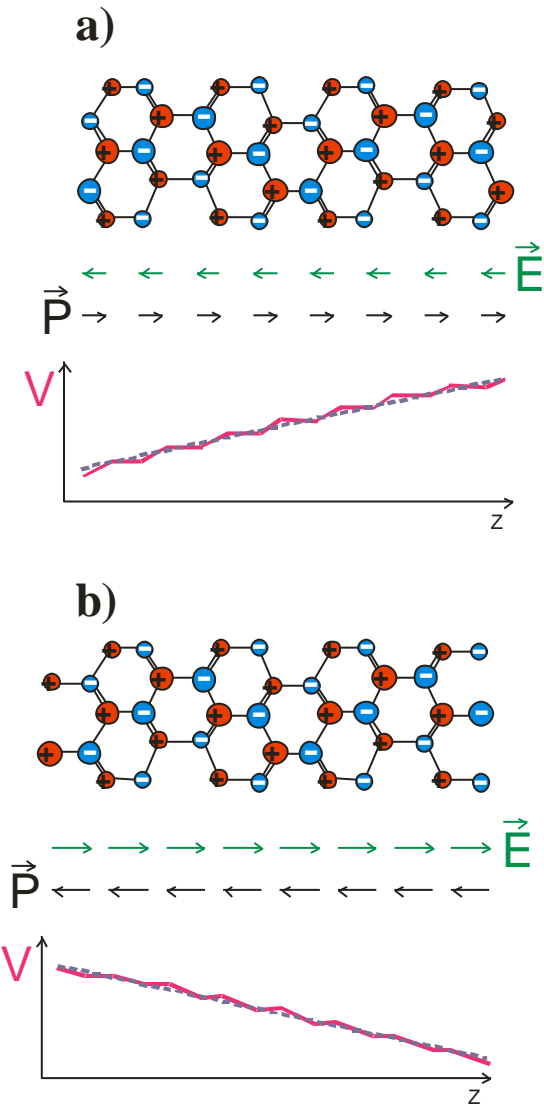

Fig. 1 Zinc blende slabs of GaN with different termination: a) by triple bonded atoms, b) by single bonded atoms. Ga and N atoms are denoted by red and blue balls, respectively. The atoms located in second layer are denoted by smaller balls. In accordance to Ref 22, it is assumed that polarization is induced by charge shift from Ga to N atoms, therefore Ga and N atoms are assumed to be positively and negatively charged. The green and black arrows represent the electric and polarization fields. The magenta solid and gray dashed lines represent layer and slab averaged electric potential profiles.

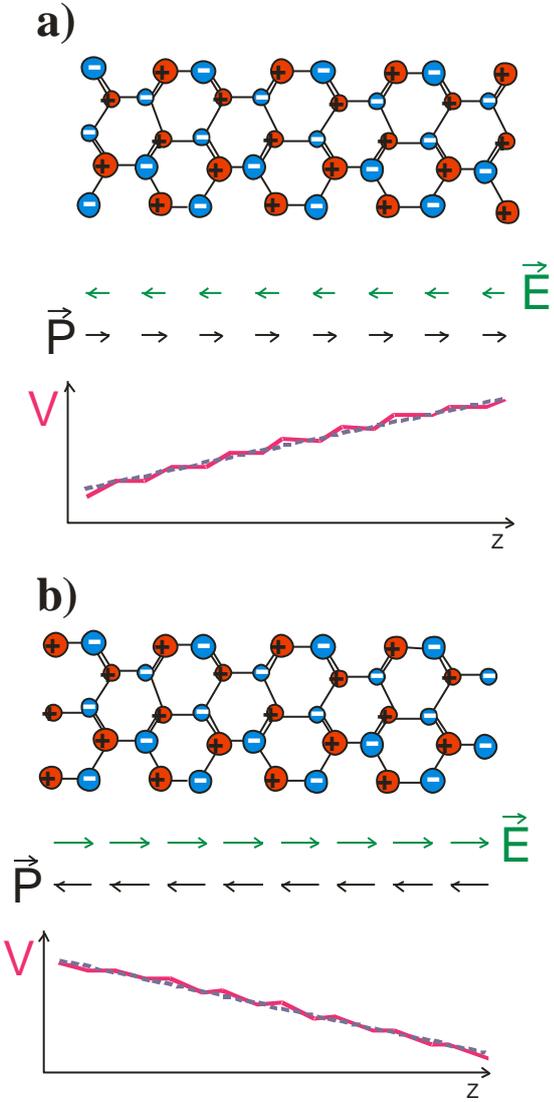

Fig. 2. Wurtzite slabs of GaN with different termination: a) by triple bonded atoms, b) by single-bonded atoms. Ga and N atoms are denoted by red and blue balls, respectively. The remaining symbols are also denoted as in Fig. 1.

From the obtained results it is evident that the polarization measured by Sawyer-Tower method, described in Ref. 22 will give nonzero results in both cases. On the other hand, the polarization in zinc blende has to vanish due to symmetry requirements. Therefore the Spaldin implementation is an oversimplified model of polarization phenomena in crystals. The misinterpreted results are obtained due to the absence of separate treatment of the two different factors: (i) the polarization that is arising due to the electron shift in the bonding of the bulk, (ii) the surface effect due to the surface charge contribution.

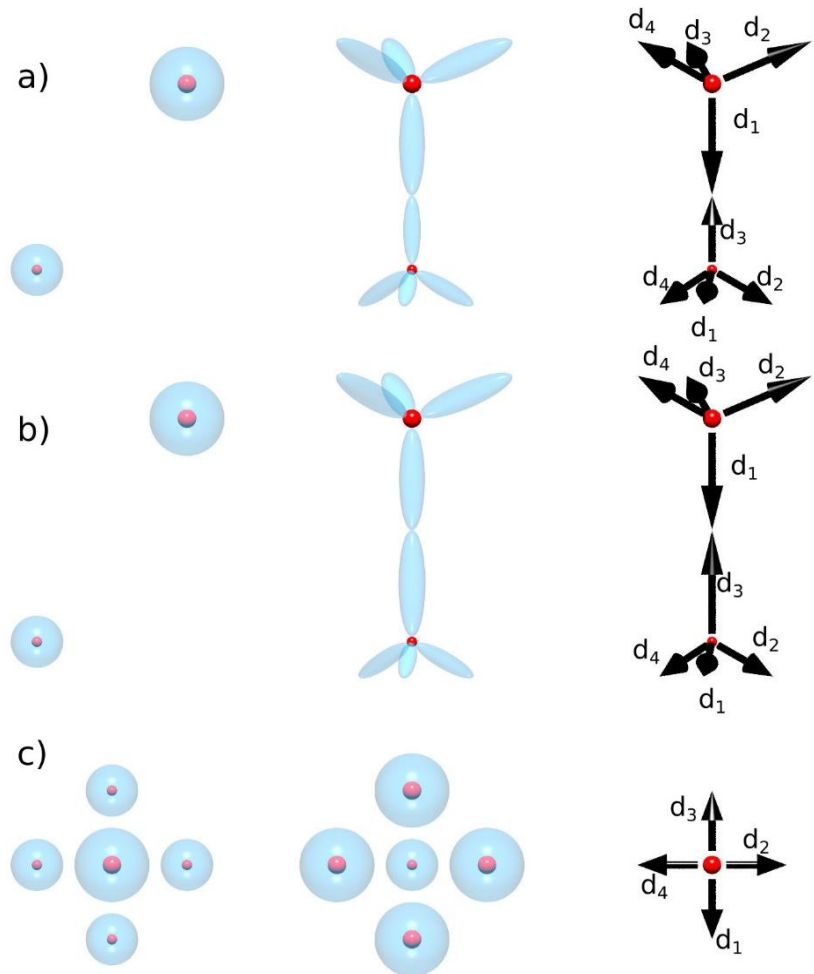

Fig. 3. Emergence of the polarization in the bonding of crystals, left - initial distribution of electronic and protonic charge, center – distribution of the charge in the bonding, right – dipole representation of the charge in the bonding. The diagrams present: a) zinc blende, b) wurtzite, c) ionic crystal. The blue and red color denotes electronic and protonic charge, respectively.

In the analysis of the first component, it is necessary to stress out that the polarization cannot be reduced to the electron transfer between atoms. In the case of the covalent bonded solid, polarization emerges due to charge transfer between crystal and atomic states in the vicinity, where the latter are understood as bonding states or, in the *ab initio* language, the valence states. Thus the correct scenario of the polarization emergence in the case of wurtzite and zinc blende lattice is presented in Fig. 3. The polarization emerges due to bonding, i.e. the bond tetrahedra should be counted. In the case of ionic compounds, this shift may be interpreted as the transition to other states, nevertheless, it is necessary to distinguish between the surface and the bulk effect. In Spaldin publication, these contributions are mixed. The proper interpretation as shown in Fig. 3 (c). In this case the polarization is zero while using Spaldin argument it is not. The difference is due to the surface effect which has to be subtracted.

The set of isolated separate atoms has polarization equal to zero, thus the spontaneous polarization is equal to the polarization change attained in the bonding. The polarization may be defined in mixed electron-proton form as:

$$\vec{P} = \frac{e}{V}\left(\int_V \vec{r}\rho_{tot}(\vec{r})d^3r\right) = \frac{e}{V}\left(\sum_{j=1}^{N} Z_j \vec{r}_j - \int_V \vec{r}\rho_{el}(\vec{r})d^3r\right) \qquad (1a)$$

where j – are the indices of all atoms, N – number of atoms, V – the volume. The mixed total charge density is:

$$\rho_{tot}(\vec{r}) = \sum_{j=1}^{N} Q_j \delta(\vec{r} - \vec{r}_j) - e\rho_{el}(\vec{r}) \qquad (1b)$$

where the charge of the nucleus of j-th atom is: $Q_j = Z_j e$, and e – is the elementary charge and the electron density, obtained from the summation over all basis functions $\varphi_q(\vec{r})$ of DFT solutions as

$$\rho_{el}(\vec{r}) = \sum_q f_q |\varphi_q(\vec{r})|^2 \qquad (1c)$$

with the occupation probability given by Fermi-Dirac distribution function with the electronic temperature $T_{el}$. This formula may be reformulated in terms of the created dipoles as:

$$\vec{P} = \frac{1}{V}\left(\sum_{j=1}^{N}\sum_{i=1}^{m}\vec{d_i}\right) \qquad (2)$$

where the first sum runs over all atoms, the second runs over all bonds, represented as dipoles (the number of dipoles is set to *m*, in the case of zinc blende and wurtzite $m = 4$). This representation assures that polarization in zinc blende vanishes while in the wurtzite it does not. Therefore this representation is fully compatible with the symmetry requirements.

It is necessary to discuss the problem of the boundaries, or more precise surface states, as the finite systems are the only ones that can exist. The boundaries entail the surface effects, i.e. charged surface states. Surface contribution is therefore defined as the difference between the crystal properties of the actual surface and the ideal continuation of the bulk. That can be represented as shown in Fig 4. Thus we apply Gibbs definition of the surface, according to which the solid bulk properties are homogeneous towards the surface, including symmetry, and the excess is associated with the surface. Therefore the symmetry of the lattice is applied to the spontaneous polarization and the surface properties are associated with the surface polarity [36]. This removes the surface dependence and related spontaneous polarization ambiguity claimed in Refs 14 – 22 [14 - 22]. On the other hand there is a method to distinguish between these two quantities. This is related to screening. In fact in any macroscopic crystal, both insulating and conductive, band screening leads to zero field in the bulk. This was identified by Meyer & Marx in their application of slab model to the simulations of ZnO polar surfaces [37]. As they show, for macroscopically large sample the band bending due to surface charge related field leads to the emergence of the mobile screening charge in conduction and valence bands, i.e. electrons and holes, respectively. The field related to polar surfaces is screened, so the quantum wells in system with no polarization should have zero field, irrespective of the polarity of the surface. The zero field result is confirmed in Fig. 8 (a) of Ref 11 showing the results of ab initio calculations of GaN/AlN MQWs with the gallium atoms located in perfect zinc blende AlN lattice, so the zinc blende symmetry is preserved [11]. On the contrary, in the case of spontaneously polarized medium such as wurtzite the field inside well and barrier emerge as illustrated in Fig. 8 (b) of Ref 11 [11].

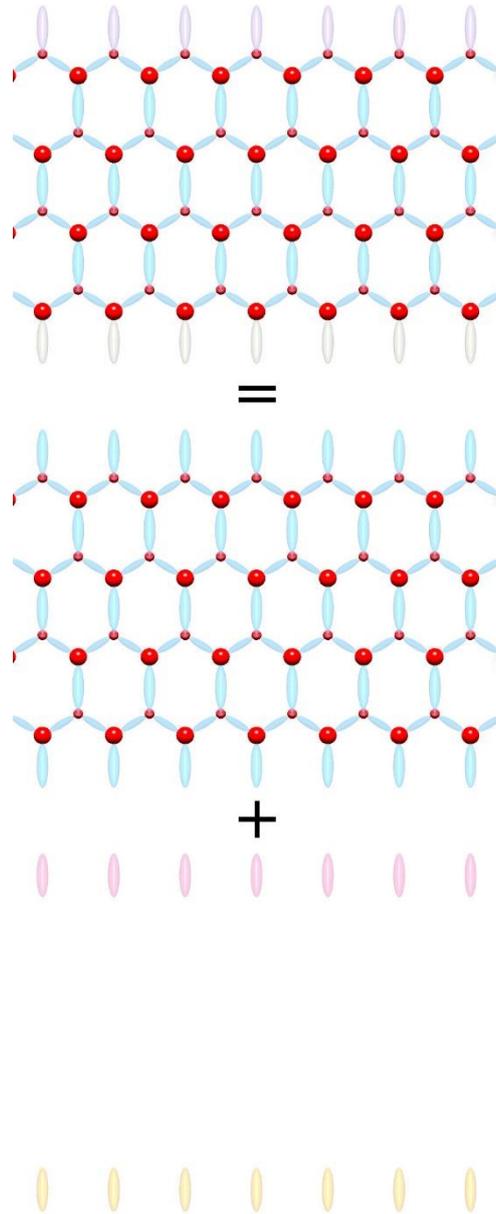

Fig. 4. Separation of the polarization and the surface states. Top - the finite slab includes contribution for both polarization and the surface (surface quantum states) at both sides, different from the bulk and also possibly form each other, center – ideal polarization system, bottom – contribution from the surface states.

In order to determine the polarization, based on the separation of the surface and bulk contributions, it is necessary to design a model capable of obtaining polarization of the bulk system without the surface. This could be done using Landau's definition with the application of the single-cell system with proper periodic boundary conditions (PBC). In Fig. 5 the two implementations of Landau definition are presented.

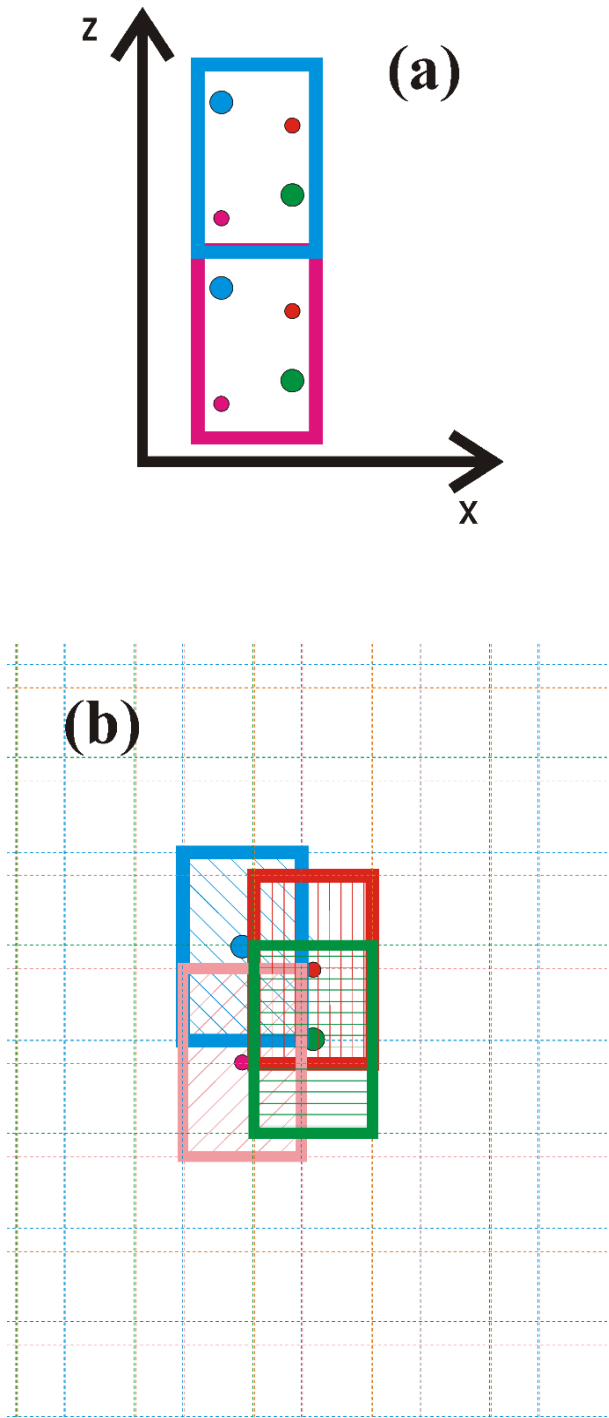

Fig. 5. Two models used for calculation of the dipole of the unit cell of the nitride semiconductor: (a) standard cell – with shift along 0z axis; (b) local atomic charge redistribution model, composed of the cells associated with the atoms. The colors denote 4 atoms: blue and green – nitrogen, red and magenta – metal (B, Al, Ga, In). The thick lines mark the cells used in the calculation of the dipole, and the dashed lines – denote the multiple copies, spanning the entire space. The color of the lines denotes association with the atoms.

The first model, presented in Fig. 5 a, is constructed from the basic simulation of periodic cells by the controlled shift of the cell along 0z direction. The obtained dipole value of the AlN wurtzite unit cell is presented in Fig. 6.

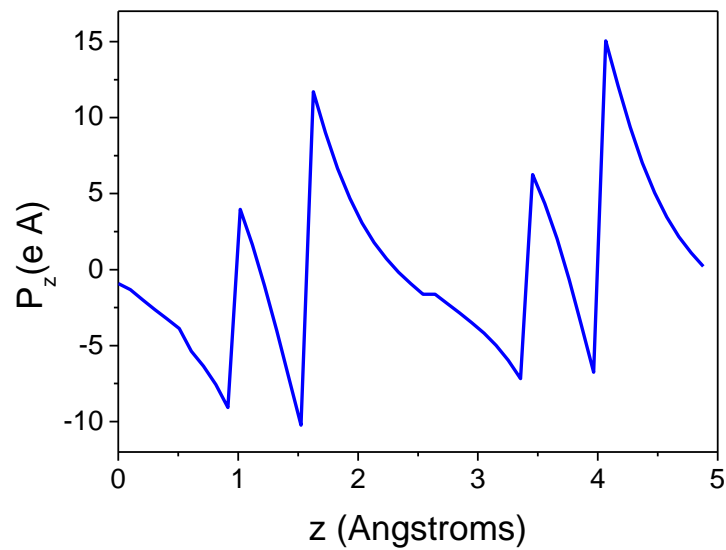

Fig. 6. Dipole moment of the AlN wurtzite unit cell in the function of the shift of the cell along the c-axis.

As it is shown, the dipole moment changes in the function of the location of the calculation cell, i.e. with the shift along the c-axis. The moment undergoes jump when the cell boundaries are crossed by Al or N atoms. In between the dipole changes continuously, proving that no single dipole value could be associated with the cell. Thus calculation of the unique value of the polarization of AlN wurtzite is not effective in this way because any value in this range is equally valid. This is in fact demonstration of the influence of the reflection symmetry breaking with respect of 0z axis by the boundaries of the calculation cell. In the proper approach the geometry of the calculation procedure (i.e. cell geometry) should be compatible with the reflection symmetry. Any symmetry breaking introducing additional components, such as the cell boundaries, leads to the incorrect value of the polarization.

A second model, presented in Fig. 5 (b) is essentially an extension of the concept of the bond creation by redistribution of electron charge, presented in Fig. 3. The dipole moment of the cell is calculated as a sum of the moments obtained for each atom (i) separately:

$$\vec{d} = \sum_{i=1}^{m} \vec{d}_i \quad (2a)$$

where the dipole related to atom (i) is calculated as:

$$\vec{d}_i = \left[ Z_i \vec{R}_i - \int_{V_i} \vec{r} \rho_{el}(\vec{r}) d^3 r \right] e \quad (2b)$$

$V_i$ – volume of the cell, centered on atom (i). Thus, any atom is surrounded by the cell which is symmetric with respect to the inversion, i.e. to reflections relative to all three axes. The overall electron density field is periodic repetition of a single calculation cell. Thus this new, atom associated cell is defined as the calculation cell of the volume $V_i$ centered on the selected atom (i). These atom associated cells have their electron charge normalized to the valence charge of specified atoms. For the atom cell (i) the calculated dipole corresponds to the emergence of the moment due to the displacement of the atom (i) charge. In summary, in the region of the overlap of four cells, denoted by different colors in Fig. 5, the density is equal to the density obtained from DFT calculations. In the other cell, the same total density is obtained from contribution from the ones marked and the neighboring repetition cells. They are marked by dashed color border lines in the entire space. Therefore the overall electron density, composed of a sum of all atom contributions is equal to that obtained from *ab initio* calculations in the entire space. At the same time this density is a patchwork of the single atoms contributions.. The total dipole moment of the cell is the sum of the moment of all atoms. As the total charge from the single atom cell (i) is electrically neutral, the obtained dipole moment does not depend on the coordinate system, therefore it could be added to the total. The entire moment is the sum of the cells dipole moments, therefore the entire moment divided by the cell volume gives the polarization of the solid. The geometry of the single cells is symmetric with respect to the reflections relative to three axes, thus not breaking the reflection symmetry.

3. **The calculation method**

The majority of the *ab initio* calculations was made using commercial Vienna Ab-initio Simulation Package (VASP) provided by University of Vienna [38-41]. This density functional theory (DFT) code uses momentum basis functional set for solution of Kohn-Sham nonlinear equations. These planar wavefunctions are marked by the momentum vector values $\vec{k}$. The maximal value of the momentum vector is determined by the energy cutoff value, which is set arbitrarily using the maximal kinetic energy cutoff value $E_{cut} = \frac{\hbar^2 k^2}{2m}$. The density of the $\vec{k}$ points is determined by the size of the system ($L_i, i = x, y, z$) by the period boundary conditions (PBC) $\left(k_i = \frac{2\pi}{L_i}\right)$. The same PBC conditions are applied for the solution of coupled Poisson equation via Fourier series. In the present solution the cutoff energy was set to $E_{cut} = 400\ eV$

The planar wavefunction set for the all electron solution of the system consisting of the metal atoms: boron, aluminum, gallium, and indium, and also nitrogen atoms is prohibitively large, thus even for a relatively small system size the reduction of the basis is required. Therefore, the electron sets of all atoms are divided into two separate classes. The first set consists of the atomic core electrons. These electrons are not considered explicitly. In fact, this set consists of closed shell electrons that are relatively affected by the crystal bonding in marginal degree only. Therefore the atomic cores are frozen with correction of the polarization effects taken into account only. The second set, considered explicitly, is denoted as valence electrons, which is equal to the total number of electrons in the simulation cell. This separation requires a special formulation in which the Coulomb potential is replaced by the procedure in which the regular function, or even the set of matrix elements is used. In VASP the norm-conserving or projector-augmented wave (PAW) potentials generated by Kresse were available [42, 43].

The standard *ab initio* method results for semiconductors are defective because they provide the energy bandgaps which are about 30% smaller than those observed experimentally. Therefore standard DFT functional is supplemented by Heyd-Scuseria-Ernzerhof (HSE) functional that is essentially an augmentation of the standard DFT functional by Hartree Fock set of equations [44]. This implementation is numerically costly, nevertheless, it is optimal for small size of the simulated systems. The experimental data are commonly used for the verification of the quality of parameterization. The lattice parameters of the bulk wurtzite boron nitride, obtained from our *ab initio* calculations are: $a_{BN}^{DFT} = 2.5417$ Å and $c_{BN}^{DFT} = 4.2019$ Å. The synthesis of wurtzite BN is extremely difficult, nevertheless the lattice parameters of w-BN were measured by x-rays giving $a_{BN}^{exp} = 2.550$ Å and $c_{BN}^{exp} = 4.227$ Å [45]. Thus for BN the *ab initio*/x-ray agreement is reasonably good. DFT lattice data for wurtzite AlN are: $a_{AlN}^{DFT} = 3.1126$ Å and $c_{AlN}^{DFT} = 4.9815$ Å. They are in reasonable agreement with the x-ray measurement data of bulk AlN wurtzite: $a_{AlN}^{exp} = 3.111$ Å and $c_{AlN}^{exp} = 4.981$ Å [46]. The calculated values for wurtzite GaN are: $a_{GaN}^{DFT} = 3.1955$ Å and $c_{GaN}^{DFT} = 5.2040$ Å, remaining in good agreement with x-ray data: $a_{GaN}^{exp} = 3.1890$ Å and $c_{GaN}^{exp} = 5.1864$ Å [47]. For InN these data are: $a_{InN}^{DFT} = 3.5705$ Å and $c_{InN}^{DFT} = 5.7418$ Å. They are in good accordance with the experimental data for wurtzite InN: $a_{InN}^{exp} = 3.5705$ Å and $c_{InN}^{exp} = 5.703$ Å [48].

HSE approximation is capable to obtain the energy bandgaps for wurtzite nitrides in general agreement with the data from optical measurements. For wurtzite boron nitride the obtained energy gap value is $E_g^{DFT}(BN) = 6.77$ eV. The experimental data for wurtzite BN are scarce, the measured bandgap is: $E_g^{exp}(BN) = 6.8\ eV$, confirming good agreement of HSE and experimental results [49]. The HSE bandgap of AlN was: $E_g^{DFT}(AlN) = 6.19\ eV$ in good agreement with the experimental data of Silveira et al. ($E_g^{exp}(AlN) = 6.09\ eV$) [50]. The *ab initio* bandgap of w-GaN was $E_g^{DFT}(GaN) = 3.41\ eV$ in agreement with $E_g^{exp}(GaN) = 3.47 eV$ [51,52]. The HSE bandgap of indium nitride was calculated to be: $E_g^{DFT}(InN) = 0.90\ eV$. The optical InN bandgap was subject of long discussion with the final consent set to $E_g^{exp}(InN) = 0.65\ eV$ [53-55].

The electron charge distribution is given as the set of the density values in the rectangular lattice points. Therefore the data are in fact a discrete representation of the continuous field. In calculation of the dipole moment, the charge can be summed first in the plane perpendicular to the dipole vector. This generates the uniaxial density distribution which is plotted in Fig. 7. The number of the density points may be controlled. The obtained density distribution is essentially symmetric, with no indication of any visible shift of the charge, also for the plot along 0z direction. This confirms that the polarization is extremely tiny effect which could be determined by extremely precise calculations.

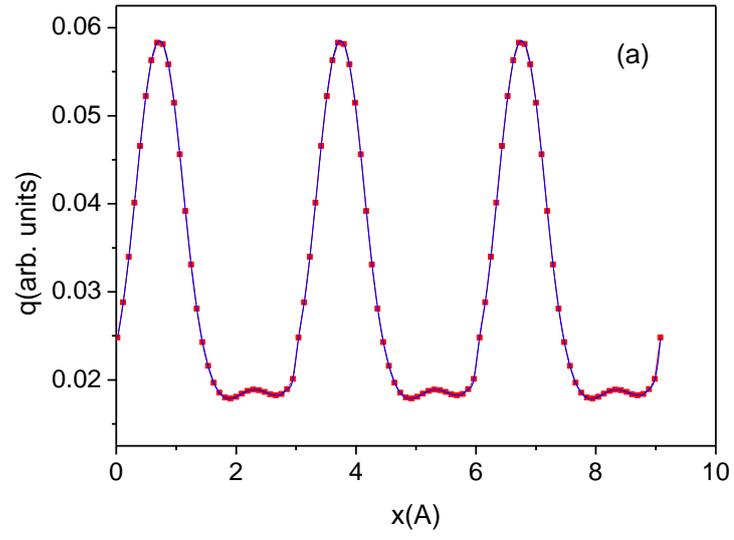

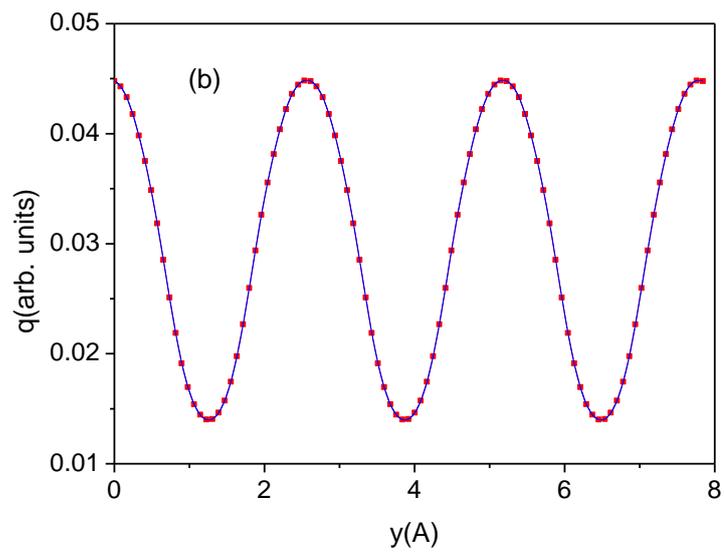

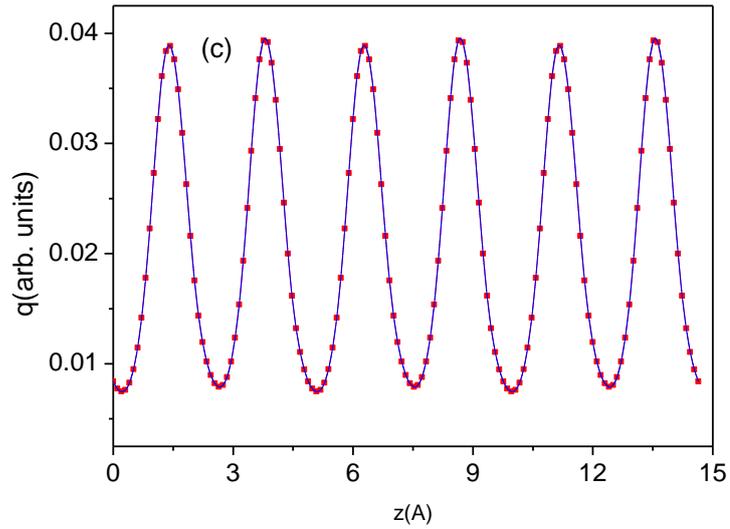

Fig. 7. Axial density distribution for wurtzite AlN. The distribution is plotted on the length of 3 lattice constants along: (a) 0x axis - over 9.338 Å , (b) 0y axis - over 8.087 Å, (c) 0z axis – over 14.944 Å. The red points represent the DFT obtained plane averaged values, the blue line is cubic spline approximation of these data.

The electron density cannot be summed directly because this generates errors that are much larger than the calculated effect. In order to mimic the smooth electron density more correctly, spline cubic functions were used. The plotted distribution is sufficient to integrate the density not only in the basic but also in the shifted cell. Since the atom-centered cell may be extended over neighboring cells, the density was calculated over three neighboring cells. Therefore the spline approximation was made over the extended distance. The plots prove that the connection between cells is smooth, correctly recovering periodic density distribution.

## 4. Results

### a. Electric dipole calculations - wurtzite

In VASP the electron density output is given as the values on the lattice of the equidistant points parallel to all basic axes. In dipole calculation, the charges could be averaged (summed) in the plane perpendicular to the dipole axis. Thus the sequence of the charge distribution along the selected axis is obtained. The examples of such distributions are plotted in Fig. 7. The number of the divisions along $0z$ axis was changed so that the coarse-grained approximation to smooth charge distribution becomes more precise for a higher number of divisions. In order to limit the computer resources needed, the number of divisions along $0x$ and $0y$ axes was not changed, equal to 33. Therefore the z-component of the dipole of the cell, and consequently the polarization along $0z$ axis depends on the number of divisions. In Fig. 8 the z-component of the polarization of wurtzite nitrides: BN, AlN, GaN and InN in function of the number of divisions of the c lattice parameter is plotted.

In the case of wurtzite structure the cell consists of 4 atoms: two N and two Me (B, Al, Ga, In) atoms. Simulation of BN employed cell of the following parameters: $a_{BN}^{DFT} = 2.5417$ Å and $c_{AlN}^{DFT} = 4.2019$ Å. The base area of the BN cell was $S_{BN}^{DFT} = 5.631$ Å², the volume $V_{BN}^{DFT} = 2.351$ Å³. Simulation of AlN employed cell of the following geometry: $a_{AlN}^{DFT} = 3.113$ Å and $c_{AlN}^{DFT} = 4.982$ Å so the base area was $S_{AlN}^{DFT} = 8.382$ Å², and the volume $V_{AlN}^{DFT} = 41.796$ Å³. In the case of GaN these data were: $a_{GaN}^{DFT} = 3.1955$ Å and $c_{GaN}^{DFT} = 5.2040$ Å and accordingly $S_{GaN}^{DFT} = 8.843$ Å² and $V_{GaN}^{DFT} = 46.020$ Å³. Finally, the InN lattice parameters were $a_{InN}^{DFT} = 3.5705$ Å and $c_{GaN}^{DFT} = 5.7418$ Å and accordingly $S_{GaN}^{DFT} = 11.041$ Å² and $V_{GaN}^{DFT} = 63.392$ Å³.

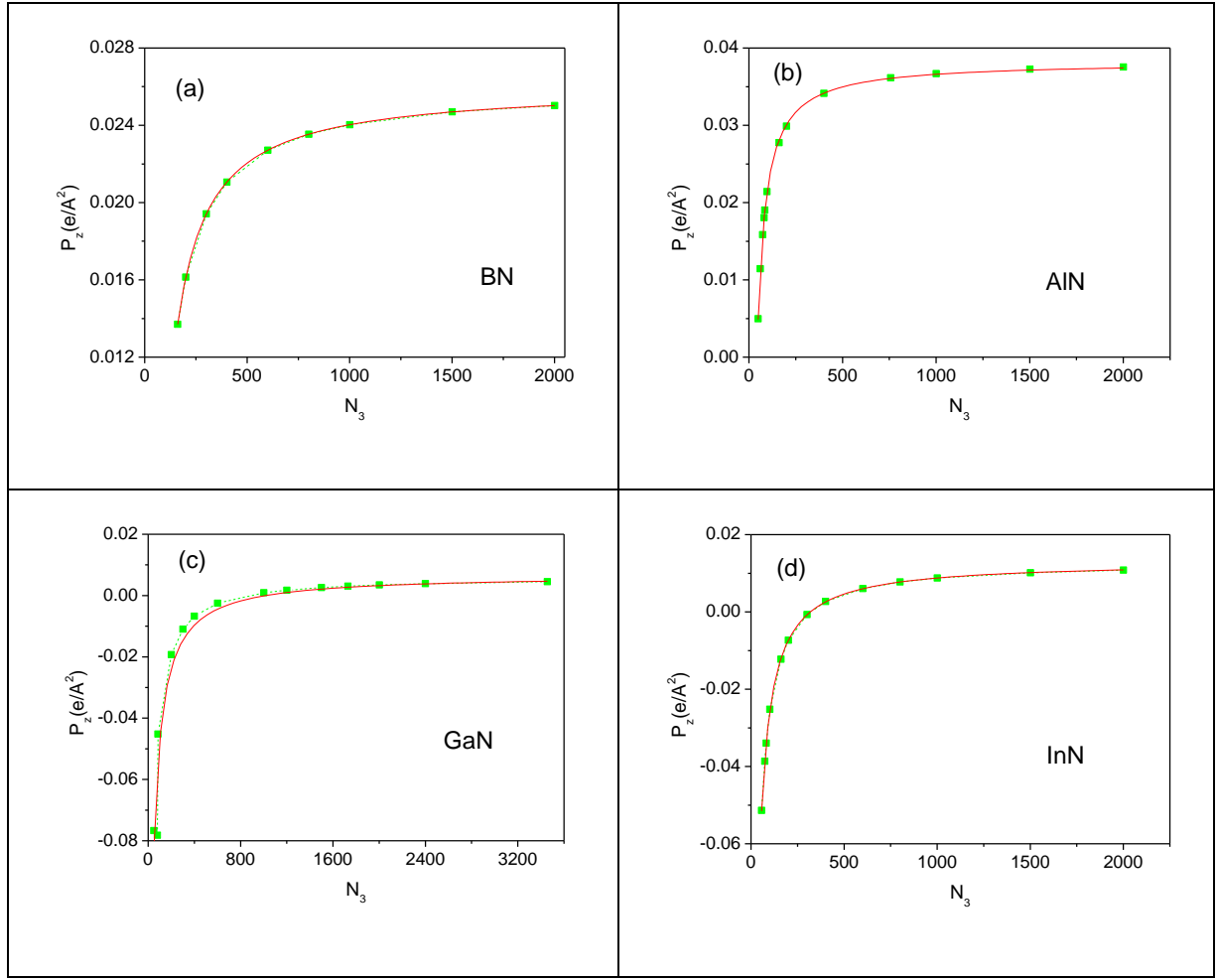

Fig. 8. Spontaneous polarization z-component $P_z$ of the wurtzite nitrides: a) BN, b) AlN, c) GaN, d) InN in the unction of the number of divisions of the cell along $0z$ axis: $N_3$. The number of divisions along two other axes was $N_1 = N_2 = 33$. The green dashed lines are for guiding the eye only, the red line is an approximation in accordance with Eqs. 3.

The simulation cell vectors for wurtzite were: $\vec{u}_1 = [a, 0,0]$, $\vec{u}_2 = [-a/2, a\sqrt{3}/2, 0]$ and $\vec{u}_3 = [0,0,c]$. Thus the divisions increase the number of points along $0z$ axis only. The relatively small number of division is partially compensated by summation in the plane perpendicular to c-axis. Nevertheless it is possible that additional systematic error is introduced.

The obtained dipole $P_z$ in the function of the number of intervals $N_3$ behaves similarly for all nitrides, and the magnitude of dipole increases to achieve final asymptotic value. In the case of GaN and InN the dipole changes sign. Thus, the asymptotic behavior of the polarization is a definite confirmation of the nonzero polarization value in all nitrides. The fit to the obtained data gives the following approximate dependence (in $e/\text{Å}^2$):

$$\vec{P}_z(BN) = 0.026 - 1.9/N_3 \tag{3a}$$

$$\vec{P}_z(AlN) = 0.038 - 1.6/N_3 \tag{3b}$$

$$\vec{P}_z(GaN) = 0.007 - 7.0/(N_3 + 29.8) \tag{3c}$$

$$\vec{P}_z(InN) = 0.013 - 4.3/(N_3 + 9.0) \tag{3d}$$

Therefore the obtained polarization values correspond to the asymptotic values for $N_3 \to \infty$.

In addition to the z-component, the polarization values in the direction perpendicular to $0z$ axis could be obtained. The stringent numerical limitations allow us to increase the number of the points along a single axis, therefore the increase in point density is possible for the case of $0y$ axis. In the case of $0x$ axis the increase for two principal axes is needed. In this way the polarization of *y*-component of AlN was calculated and the results are presented in Fig. 9. As it is shown, the polarization component is much lower, $\vec{P}_y(AlN) \cong 2.7 \times 10^{-4}~e/\text{Å}^2$. It is not zero nevertheless this is extremely low value, not giving a nonzero value but showing precision of the representation of the density field. Thus any finite number of points cannot give the value of the polarization below some limit, in our case that was $\Delta P \sim \vec{P}_y(AlN) \cong 2.7 \times 10^{-4}~e/\text{Å}^2$. In fact this is in agreement with the zero polarization value which is in accordance with the symmetry requirements.

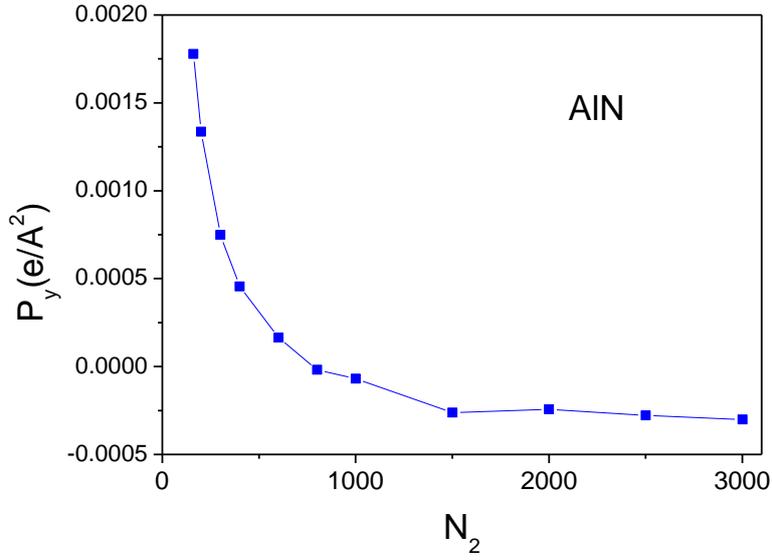

Fig. 9. Spontaneous polarization y-component $P_y$ of wurtzite AlN in function of the number of division of the cell length along $0y$ axis: $N_2$. The number of divisions along two other axes were $N_1 = 33$ and $N_3 = 49$.

### b. Electric dipole calculations – zinc blende

Additional verification of the basic model stems from the calculation of zinc blende polarization values of these nitrides. According to symmetry argument, the polarization is zero. The lattice constant of AlN was $a_{AlN-zb}^{DFT} = 2.680$ Å. The calculation cell of the volume $V_{AlN-zb}^{DFT} = 31.421$ Å$^3$ contains 6 atoms: 3 Al and 3N. The simulation cell vectors for zinc blende were $\vec{u}_1 = [a, 0, 0]$, $\vec{u}_2 = [-a/2, a\sqrt{3}/2, 0]$ and $\vec{u}_3 = [0, 0, c]$. Thus, the convenient calculation was possible for z-component only. The calculated result for polarization of zinc blende AlN is presented in Fig. 10. As it is shown, the polarization values are relatively high, but they are decreasing continuously. This is related to the fact that the density of lattice in the plane perpendicular to triple axis is small and some values in the plane perpendicular, compensating the bond parallel to triple axis are missing. Thus the error in the determination is large. Nevertheless these data are consistent with the zero value of spontaneous polarization in zinc blende crystals, in accordance with the symmetry arguments.

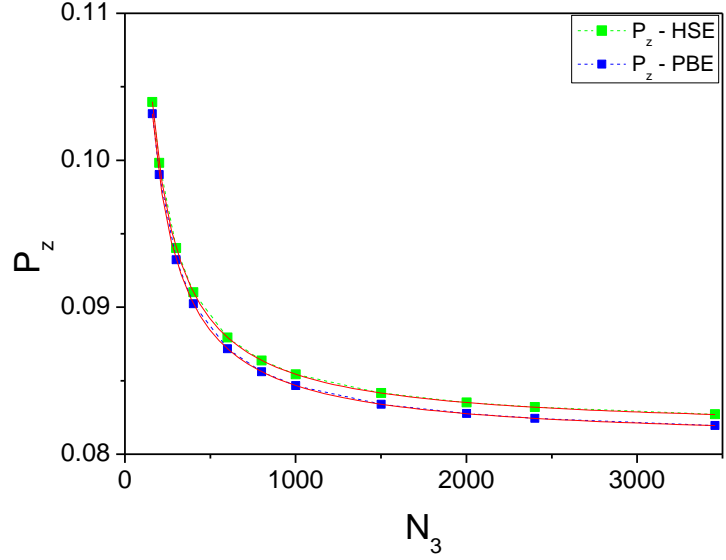

Fig. 10. Spontaneous polarization z-component $P_z$ of the zinc blende AlN in the function of the number of division of the cell length along $0z$ axis: $N_3$. The number of divisions along two other axes were $N_1 = N_2 = 33$. The green and blue symbols denote data obtained for HSE and PBE approximations, respectively. The dashed lines are for guiding the eye, the red solid lines are approximations in accordance with Eqs 4.

These data indicate the monotonous decrease of the polarization values for increased number of intervals. The following approximations for these data were obtained:

$$\vec{P}_z(HSE) = 0.0816 + 3.99/(13.85 + N_3) \qquad (4a)$$

$$\vec{P}_z(PBE) = 0.0808 + 3.92/(13.72 + N_3) \qquad (4b)$$

These data prove that both approximations give essentially identical values of polarization. The difference is minor. On the other hand, the asymptotic value is not zero which is related to the small number of points in the perpendicular plane so the cancellation of the three dipoles at angle with that along the c-axis is not complete. On the other hand, the data for wurtzite indicated a monotonous increase of the dipole magnitude while for zinc blende is the opposite. This again confirms the disappearance of the polarization in the latter case.

### c. Spontaneous polarization and zero field polarization (Berry phase).

The above values were obtained for the zero electric field, as this is the only condition compatible with the solution of Poisson equation using the fast Fourier transform (FFT) method. This is different from spontaneous polarization state which is defined as the emergence of dipole moment and the electric field without any external contribution. Therefore, the latter assumes the zero value of the electric displacement field in the entire system, i.e. $\vec{D} = 0$ [31]. That assumption determines the relation between the spontaneous polarization $\vec{P}_{i,s} = \vec{P}_o$ and electric field $\vec{E}_{i,s}$ inside the polarized medium (the indices denote: i – internal, s – spontaneous):

$$\vec{P}_{i,s} = \vec{P}_o = -\varepsilon_o \vec{E}_{i,s} \qquad (5)$$

Electric dipole vector is directed from negative to positive charge while the electric field is the force acting on positive charge i.e. it is directed opposite. Assume that we consider an infinite polar slab. Then the electric field related to spontaneous polarization outside the slab $\vec{E}_{e,s}$ (index e – denotes external, i – internal ) vanishes, i.e. $\vec{E}_{e,s} = \vec{D} = 0$ . In the calculation of spontaneous polarization employing Berry phase formulation [25, 26] Resta assumed that the electric field vanishes, i.e. $\vec{E}_{i,B} = 0$ (B – denotes Berry state). From the spontaneous condition $\vec{D} = 0$ it follows that Berry phase polarization $\vec{P}_{i,B}$ should vanish, i.e. $\vec{P}_{i,B} = 0$ . This is not the case, therefore the condition $\vec{D} = 0$ is not fulfilled in the Berry state, thus this state requires nonzero external electric field $\Delta \vec{E}_{iB}$ to be added to the spontaneous polarization field so that $\Delta \vec{E}_{iB} + \vec{E}_{i,s} = 0$ . This additional field obeys the linear regime with the continuity of electric displacement field $\vec{D}_i = \vec{D}_i$ therefore the external compensating field $\Delta \vec{E}_{eB}$ is:

$$\Delta \vec{E}_{eB} = \varepsilon \, \Delta \vec{E}_{iB} = \frac{\vec{P}_o \, \varepsilon}{\varepsilon_o} \qquad (6)$$

where $\epsilon$ is dielectric permittivity. The application of this field induces the polarization change $\Delta \vec{P}$:

$$\Delta \vec{P} = \varepsilon_o \chi \Delta \vec{E}_{iB} = \chi \vec{P}_o \qquad (7)$$

where the dielectric susceptibility is: $\chi = \epsilon - 1$. Since the Berry polarization $\vec{P}_B$ is the sum of the spontaneous polarization $\vec{P}_o$ and polarization change $\Delta \vec{P}$, i.e. $\vec{P}_B = \vec{P}_o + \Delta \vec{P}$ we obtain the final result:

$$\vec{P}_B = \epsilon \vec{P}_o \qquad (8)$$

Thus the Berry (zero field) polarization is different from spontaneous polarization by the factor equal to the dielectric permittivity of the material. Therefore the data from Berry (zero field) polarization determined above may be used to determine the spontaneous polarization.

### d. Multiquantum well (MQWs) / superlattice calculations - wurtzite

Additional verification of the polarization values may be obtained indirectly from *ab initio* calculations of polar GaN/AlN, InN/GaN, BN/AlN and InN/AlN superlattices, which are utilized as multiquantum wells (MQWs) in optoelectronic devices. These structures are very thin, therefore the polarization-induced electric fields are not screened giving rise to quantum Confined Stark Effect (QCSE) [8-12]. In most cases, the Fermi level in the bulk semiconductor, on both sides of the structure, is pinned by the same defect. Thus the Fermi level position, and accordingly the potential difference is approximately zero over entire well-barrier system, so that the electric fields in the well ($E_w$) and in the barrier ($E_b$) are proportional to polarization difference only [9-11,30]. Therefore it is assumed that potential is periodic with respect to a single well-barrier length that can be used to derive the electric field in the wells $E_w$, and in the barriers $E_b$, as:

$$E_w = \frac{b(P_w - P_b)}{\varepsilon_o(w\varepsilon_b + b\varepsilon_w)} \qquad (9a)$$

$$E_b = \frac{w(P_b - P_w)}{\varepsilon_o(w\varepsilon_b + b\varepsilon_w)},  \quad (9b)$$

where *w* and *b* are thicknesses of the well and barrier, respectively, $\varepsilon_w$ and $\varepsilon_b$ are dielectric constants of the well and barrier, and $\varepsilon_o$ is permittivity of the vacuum. In these equations it was assumed that the potential jumps [56,9], due to dipole layers at heterointerfaces cancel out. These fields may be used to obtain the polarization difference:

$$\Delta P = P_w - P_b = \frac{\varepsilon_o(w\varepsilon_b + b\varepsilon_w)E_w}{b} = -\frac{\varepsilon_o(w\varepsilon_b + b\varepsilon_w)E_b}{w} \quad (10)$$

Such wurtzite structures were calculated using the ideal lattice positions of BN, AlN, GaN and InN lattice. The model was created such that the metal atoms are located in the sites of single nitride semiconductor lattice, e.g. Ga atoms are located in AlN lattice. No relaxation was allowed, so the layers are lattice strained. In such case the lattice is a pure single wurtzite semiconductor. Since we use an identical number of both metal layers therefore in these simulations the thickness of the well and barrier are identical, i.e. $w = b$. From these results, the fields in the wells and barriers were obtained by a linear fit to potential profiles as shown in Fig. 10.

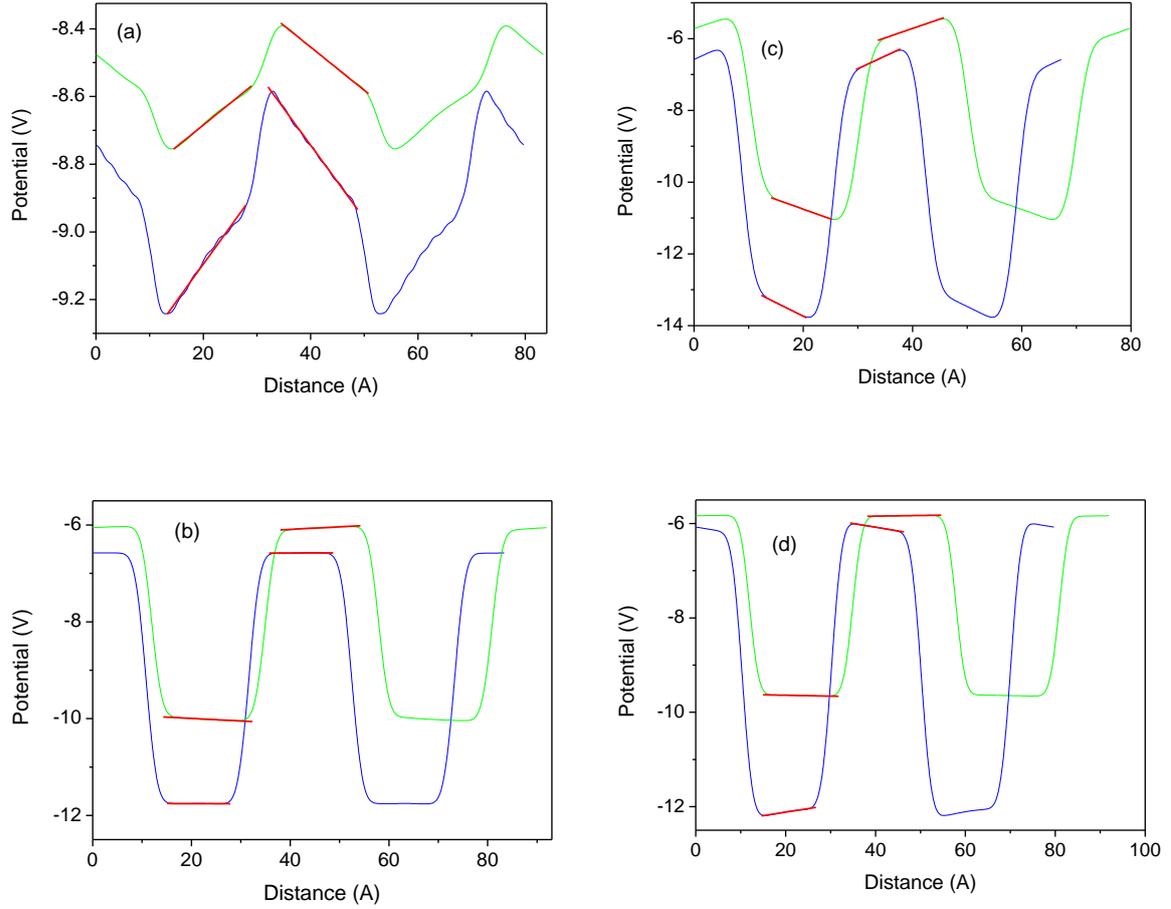

Fig. 11. Averaged electric potential profiles along $0z$ axis in wurtzite superlattice vs. distance measured in metal atomic layers (AL) determined for structures with 8 AL for both the well and the barrier thicknesses (i.e. $b = w$) : (a) AlN/GaN , (b) GaN/InN; (c) BN/AlN; (d) AlN/InN. Green and blue lines correspond to larger/smaller lattice parameters (i.e. fractionally strained/compressed), respectively. Red lines represent linear slopes of the potential, i.e. electric fields.

In fact the obtained polarization is strongly affected by the piezoelectric effects. Thus in the strained lattice the z component of the polarization is:

$$P_3 = P_{3,0} + \epsilon_{311}\varepsilon_{11} + \epsilon_{333}\varepsilon_{33} \quad (11)$$

where $P_{3,0}$ is the z-component of the spontaneous polarization $\vec{P}_o$, $\epsilon_{311}$ and $\epsilon_{333}$ are piezo constants and $\varepsilon_{11}$ and $\varepsilon_{33}$ strain tensor components. The properties of strained well/barrier systems, obtained in these simulations are summarized in the Table 2.

Table 1. The properties of lattice strained well/barrier systems

| System | Lattice | $\varepsilon_{11}(w)$ $\varepsilon_{33}(w)$ | $\varepsilon_{11}(b)$ $\varepsilon_{11}(b)$ | $E_w(V/Å)$ | $E_b(V/Å)$ | $\Delta P_l(C/m^2)$ |
|---|---|---|---|---|---|---|
| AlN/GaN | AlN | -0.0259 -0.0428 | 0 0 | 0.0212 | $-0.0217$ | 0.040 |
| AlN/GaN | GaN | 0 0 | 0.0266 0.0446 | 0.0129 | $-0.0128$ | 0.023 |
| GaN/InN | GaN | -0.1050 -0.9366 | 0 0 | $-3.52 \times 10^{-4}$ | $4.64 \times 10^{-4}$ | $8.99 \times 10^{-4}$ |
| GaN/InN | InN | 0 0 | 0.1173 0.1033 | $-0.00535$ | 0.00526 | 0.012 |
| BN/AlN | BN | -01834 0.1565 | 0 0 | 0.0706 | $-0.0759$ | 0.081 |
| BN/AlN | AlN | 0 0 | 0.2246 0.1855 | 0.0529 | $-0.0523$ | 0.111 |
| AlN/InN | AlN | -0.2594 -0.1324 | 0 0 | 0.0150 | $-0.0158$ | $3.57 \times 10^{-3}$ |
| AlN/InN | InN | 0 0 | 0.1282 0.1324 | $-0.00188$ | 0.00136 | 0.040 |

In this table, the strain component for the native lattice is zero, the second layer is assumed to be strained according to the lattice parameter difference, i.e. $\varepsilon_{\alpha\alpha} = \frac{a_{i,\alpha\alpha} - a_{j,\alpha\alpha}}{a_{j,\alpha\alpha}}$ where i,j denote well and barrier (w,b) and $\alpha = 1,3$ (coordinates), respectively.

In summary, the polarization in the strained systems has two components: spontaneous and piezo. The piezo has two contributions, related to the enforced strain along c-axis and in the perpendicular plane, i.e. the strain tensor components $\varepsilon_{11}$ and $\varepsilon_{33}$, respectively. These data are not sufficient to calculate both piezo constants as we have a single equation from these data. Therefore the second system was devised such that the strain in the plane is identical, i.e. lattice compatible, but the layers are not strained along c-axis, therefore $\varepsilon_{33} = 0$. Thus these systems are plane strained. The *ab initio* calculated electric fields along c-axis of these superlattices are presented in Fig. 12.

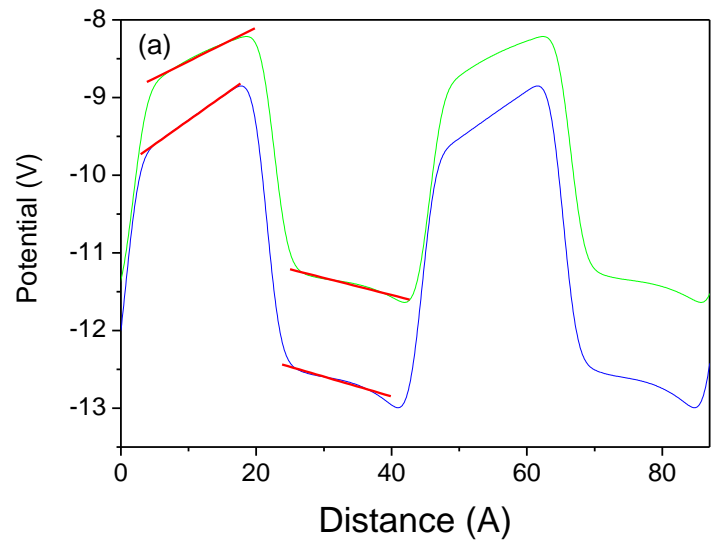

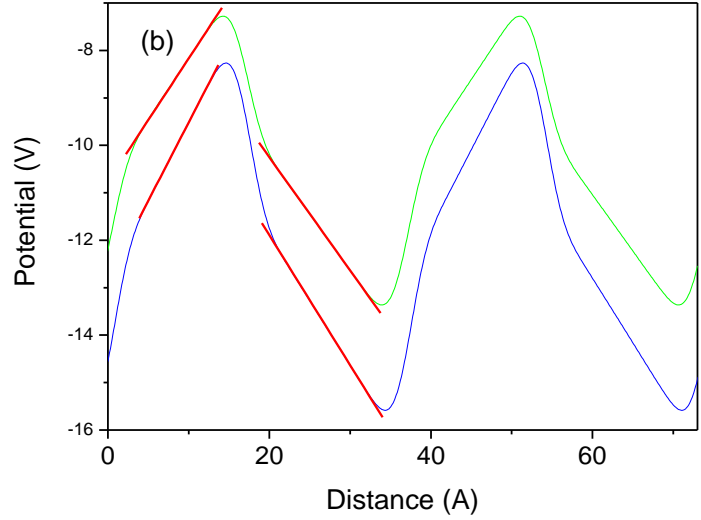

Fig. 12. Averaged electric potential profiles along 0z axis in wurtzite superlattice with the thickness of 8 metal atom layers (AL) for both the well and the barrier: (a) GaN/InN; (c) BN/AlN;. The system was strained in a plane, while it was relaxed along c-axis. Green and blue lines correspond to plane strained (i.e. $b > w$) or plane compressed (i.e. $b < w$) lattices, respectively. Red lines represent linear slopes of the potential, i.e. electric fields.

These polarization values can be used for the determination of the piezo constants. The polarization difference in lattice strained system is:

$$\Delta P_l = \Delta P_0 + \epsilon_{311}\varepsilon_{11} + \epsilon_{333}\varepsilon_{33} \qquad (12)$$

where $\Delta P_0$ is the spontaneous polarization difference of the well and the barrier, $\varepsilon_{11}$ and $\varepsilon_{33}$ are the strain components of the strained layer (the second layer is not strained so the strain component are zero), $\epsilon_{311}$ and $\epsilon_{333}$ are piezo constants of the strained layer. In case of plane strained system, the strain z-component is zero, therefore the polarization difference is:

$$\Delta P_p = \Delta P_0 + \epsilon_{311}\varepsilon_{11} \qquad (13)$$

From this set of data, the first piezo component can be obtained as:

$$\epsilon_{311} = (\Delta P_p - \Delta P_0)/\varepsilon_{11} \qquad (14)$$

and the second as

$$\epsilon_{333} = (\Delta P_l - \Delta P_0 - \Delta P_p)/\varepsilon_{33} \qquad (15)$$

The data for the plane-strained superlattices are presented in Table 2. Using the data from Tables 1 and 2 with the application of Eqs 14 and 15 the piezo constants were obtained which are shown in Table 3.

Table 2. The properties of the plane strained (zero strain along c-axis) well/barrier systems

| System | Strained | $\varepsilon_{11}(w)$ $\varepsilon_{33}(w)$ | $\varepsilon_{11}(b)$ $\varepsilon_{33}(b)$ | $E_w(V/Å)$ | $E_b(V/Å)$ | $\Delta P_p(C/m^2)$ |
|---|---|---|---|---|---|---|
| GaN/InN | InN | -0.1050 0 | 0 0 | 0.0681 | −0.0255 | 0.100 |
| GaN/InN | GaN | 0 0 | 0.1173 0 | 0.0434 | −0.0221 | 0.0705 |
| BN/AlN | AlN | -01834 0 | 0 0 | −0.2741 | 0.3269 | 0.449 |
| BN/AlN | BN | 0 0 | 0.2246 0 | −0.2384 | 0.2595 | 0.375 |

In this table, the strain component for the native lattice is zero, the second layer is assumed to be strained in plane perpendicular to c-axis, according to the lattice parameter difference, i.e. $\varepsilon_{33} = \frac{c_{i,33} - c_{j,33}}{a_{j,33}}$ where i,j denote well and barrier (w,b), respectively.

e. **Multiquantum well (MQWs) / superlattice calculations – zinc blende**

Similar calculations were made for zinc blende superlattice of the nitrides. In order to obtain elongated profiles, PBE approximation was used. As shown previously, both HSE and PBE approximations provide identical polarization values. The potential profiles are shown in Fig. 13. These data indicate that the electric fields in both cases for GaN well and AlN barrier are extremely small, thus confirming the absence of the polarization-induced fields in the zinc blende lattice.

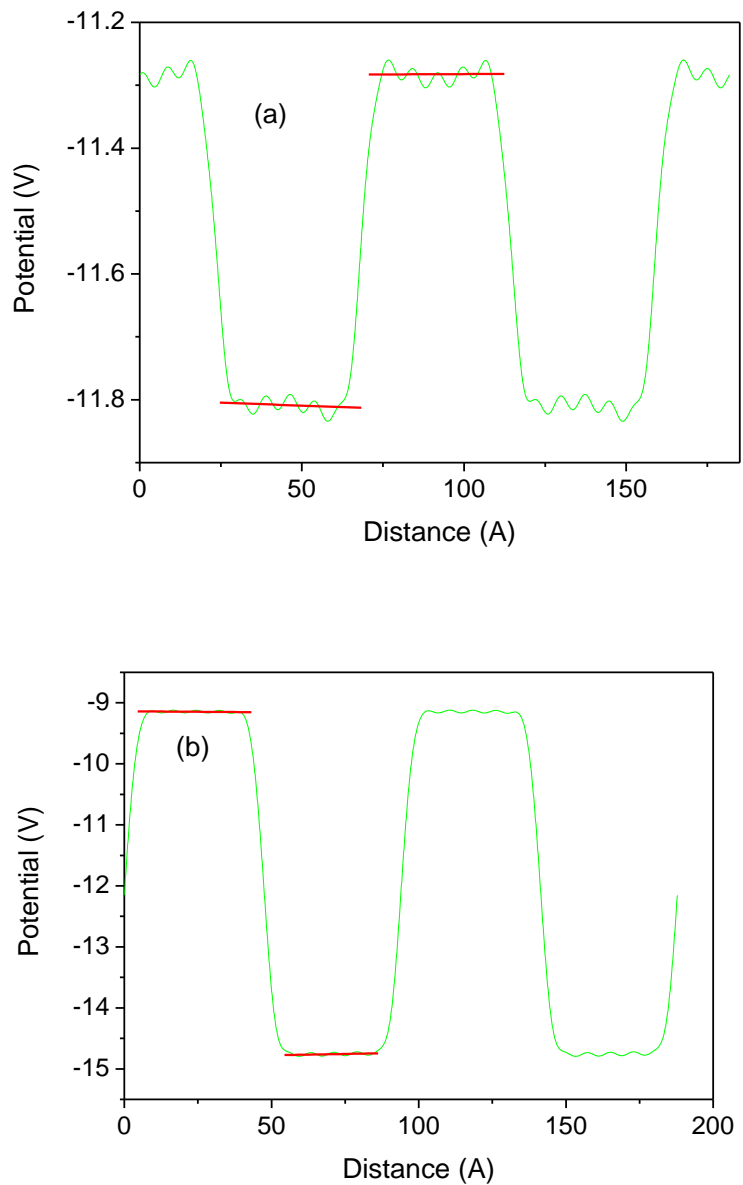

Fig. 13. Averaged electric potential profiles along [111] direction in zinc blende superlattice: (a) AlN/GaN , (b) GaN/InN. Green and blue lines correspond to larger/smaller lattice parameters, respectively.

From the linear approximation the following fields were obtained: AlN/GaN - $E_w = 2.151 \times 10^{-5}$ V/Å and $E_b = -1.897 \times 10^{-4}$ V/Å, GaN/InN - $E_w = 3.515 \times 10^{-4}$ V/Å and $E_b = 8.687 \times 10^{-4}$ V/Å. Therefore it may be concluded that these data indicate the absence of polarization induced fields in this structure.

### f. Critical comparison of the results

The dielectric permittivity of the nitrides can be determined using Green function formulation, but this approximation is burdened by relatively high error. Therefore experimental data was used: for wz-BN: $\varepsilon_{wz-BN} = 6.85$, wz-AlN: $\varepsilon_{wz-AlN} = 10.31$, wz-GaN $\varepsilon_{wz-GaN} = 10.28$ and wz-InN $\varepsilon_{wz-InN} = 14.61$. These data were used to determine the polarization values listed in Table 1. For the comparison, the data obtained in Refs. 28, 29 and 9 as compiled in Ref. 32 are listed.

Table 3. Polarization (in $C/m^2$) and piezoelectric constants of the nitrides

| Property | Ref | BN | AlN | GaN | InN |
|---|---|---|---|---|---|
| Spontaneous polarization, $P_3$ | This work | 0.061 | 0.059 | 0.011 | 0.014 |
| | [27] | | 0.081 | 0.029 | 0.032 |
| | [28] | | 0.090 | 0.034 | 0.042 |
| | [29] | | 1.351 | 1.312 | 1.026 |
| | [11] | | 0.090 | 0.019 | 0.028 |
| Piezo constant $\epsilon_{311}$ | This work | -1.17 | -0.99 | -0.64 | -0.83 |
|  | [27] |  | -0.60 | -0.49 | -0.57 |
|  | [28] |  | -0.53 | -0.34 | -0.41 |
|  | [29] |  | -0.676 | -0.551 | -0.604 |
| Piezo constant $\varepsilon_{333}$ | This work | 1.88 | 1.18 | 0.74 | 0.96 |
|  | [27] |  | 1.46 | 0.73 | 0.97 |
|  | [28] |  | 1.50 | 0.67 | 0.81 |
|  | [29] |  | 1.569 | 1.020 | 1.328 |

These polarization values are in general agreement with those obtained in Ref. 27, 28 and 11. They are much different from those obtained in Ref 29. The piezo constants are of the same order, nevertheless the values are quite different. The piezo values in the present work could be potentially affected by additional charges at the heterostructures. Therefore much larger effort should be made in the future to derive reliable piezo parameters of wurtzite semiconductors. One of the most promising is the application of the model presented in this work.

## 5. Summary and conclusions

A new way of summarizing the results obtained in this work will be used, following the following basic scheme: (i) state of art before the publication, (ii) the results of the present work (iii) state of art after publication.

The state of the art before the publication may be summarized as follows:

(a) Application of Landau model to infinite solids is questioned, and separation of surface and polarization effects is claimed to be impossible [15 – 22].

(b) Spontaneous polarization as bulk quantity was redefined in accordance with the arguments used in Berry phase formalism [25,26].

(c) Berry phase calculations of spontaneous polarization of the nitrides provided drastically different data [27 -29].

(d) Slab calculation provided data [11] that are basically compatible with several Berry phase results [27,28], but not with the second set [29,12]. Nevertheless the difference is considerable which may be also attributed to surface charge in the slab model [11]. Therefore the slab results cannot be treated as final.

(e) Superlattice calculations provide data on polarization difference that is in basic agreement with all results, both in Berry phase and slab results [32].

The results presented in this publication may be summarized in the following way

(a) Polarization as a bulk quantity of the infinite solid was redefined, and separation into spontaneous polarization and surface effects was proposed.

(b) The geometric model allowing calculation of spontaneous polarization as electric dipole density is formulated (based on Landau definition)

(c) It was demonstrated that some, earlier proposed models of the polarization provide incorrect picture of the phenomenon, mixing polarization and polar surface effects.

(d) The spontaneous polarization of wurtzite nitrides was calculated, showing that the c-axis component $P_z$ is nonzero and the others, $P_x$ and $P_y$ are zero.

(e) The calculated spontaneous polarization of zinc blende nitrides is zero.

(f) The obtained polarization values $P_z$ of wurtzite nitrides are in general agreement with the Berry phase results of Bernardini et al. [27,29] and are different from Dreyer et al. [29].

(g) The obtained polarization values $P_z$ of wurtzite nitrides are in agreement with those derived from superlattice calculations [32].

The state of art after publication can be described as:

a) Spontaneous polarization of infinite solid is defined as bulk quantity.

b) Spontaneous polarization may be calculated using Landau formulation using a geometric model.

c) Polar surfaces are different objects, independent of the spontaneous polarization.

d) Spontaneous polarization of the nitrides is further verified, showing basic agreement for all wurtzite nitrides.

e) Piezoelectric effects for wurtzite nitrides are correctly obtained.

In conclusion, it is stated that the present work made considerable progress in the basic understanding of the spontaneous polarization of the infinite solids as basic property, the polar surface as different objects and also of the determination of the parameter values of the wurtzite and zinc blende nitride semiconductors. Still, the values cannot be treated as definitely determined due to the deficiencies of the computational resources which affected the precision of the obtained data.

**Acknowledgements**

The research was partially supported by Poland National Centre for Research and Development [grant number: TECHMATSTRATEG-III/0003/2019-00] and partially by Japan JST CREST [grant number JPMJCR16N2] and by JSPS KAKENHI [grant number JP16H06418]. This research was carried out with the support of the Interdisciplinary Centre for Mathematical and Computational Modelling at the University of Warsaw (ICM UW) under grants no GB77-29, GB84-23 and GB96-1851.